# Correlated local bending of DNA double helix and its effect on the cyclization of short DNA fragments

Xinliang Xu[1, 2], Beng Joo Reginald Thio[2], Jianshu Cao[1, 3, §]

[1]Department of Chemistry, MIT, Cambridge, MA 02139, USA

[2]Pillar of Engineering Product Development, Singapore University of Technology and Design, Singapore

[3]Singapore-MIT Alliance for Research and Technology (SMART), Singapore

[§]To whom correspondence should be addressed. Email: jianshu@mit.edu (J.C.)

**Abstract**

We report a theoretical study of DNA flexibility and quantitatively predict the ring closure probability as a function of DNA contour length. Recent experimental studies show that the flexibility of short DNA fragments (as compared to the persistence length of DNA $l_P \sim 150$ base pairs) cannot be described by the traditional worm-like chain (WLC) model, e.g., the observed ring closure probability is much higher than predicted. To explain these observations, DNA flexibility is investigated with explicit considerations of a new length scale $l_D \sim 10$ base pairs, over which DNA local bend angles are correlated. In this correlated worm-like chain (C-WLC) model, a finite length correction term is analytically derived and the persistence length is found to be contour length dependent. While our model reduces to the traditional worm-like chain model when treating long DNA at length scales much larger than $l_P$, it predicts that DNA becomes much more flexible at shorter sizes, which helps explain recent cyclization measurements of short DNA fragments around 100 base pairs.

## I. Introduction

The flexibility of DNA has great impacts on its overall shape as well as on many of its biological functions, such as chromosomal DNA packaging[1], DNA damage repair[2] and regulation of gene expression[3]. Since the early 1990s many experimental techniques[4] have been developed to help us understand the mechanical characteristics of the molecule. These results have supported the so-called worm-like chain (WLC) model[5], where all structural information of DNA is coarse grained into one single fitting parameter—the persistence length $l_P \sim 150$ base pairs (bps)—under physiological conditions[6] as well as in a hydrodynamic flow field[7]. This simple model captures the essential physics underlying many mechanical properties of DNA[8, 9] and reproduces the experimental force-extension curve with impressive precision for DNA with contour length $L$ much larger than $l_P$[10]. However, recent advances in experimental techniques have provided new data[11], challenging the application of this model to short DNA fragments with $L < l_P$, and suggesting a length dependent DNA flexibility[12]. Furthermore, recent experimental studies of short DNA fragments show significant "softening" of DNA as the experimentally observed ring closure probability is orders of magnitudes higher than the WLC prediction[13]. A more comprehensive model incorporating structural details of DNA underlying these findings is needed.

The development of the WLC model starts from a discrete description of DNA, the Kratky-Porod (KP) model[14], where DNA is simplified as a succession of equal-sized segments with length $l_0$ and orientation $\vec{t}_n$ for the $n^{th}$ segment (figure 1A). Let $\theta_n$ denote the bend angle between the $n^{th}$ segment and the $(n+1)^{th}$ segment with $\cos\theta_n = \vec{t}_n \cdot \vec{t}_{n+1}$. In the most naïve random coil description, these segments are considered as totally uncorrelated, i.e. $\langle \vec{t}_i \cdot \vec{t}_j \rangle = \delta_{ij}$ where $\delta_{ij}$ is the Kronecker delta function. As an improved description, the KP model assumes that the chain resists to bending deformation, characterized by a classical elastic energy of bending defined through

$$E_{KP} = \frac{B}{l_0}\sum_{n=1}^{N-1}(1 - \vec{t}_n \cdot \vec{t}_{n+1}) = \frac{B}{l_0}\sum_{n=1}^{N-1}(1 - \cos\theta_n), \tag{1}$$

where $B = l_P k_B T$ is the bending modulus. As a result, the orientations of different chain segments are now correlated such that $\langle \vec{t}_i \cdot \vec{t}_j \rangle = e^{-l_0|i-j|/l_P}$. The WLC model can be obtained by taking the continuous limit ($l_0 \to 0$, $N \to \infty$ while $Nl_0 = L$) of eq. 1 as

$$E_{WLC} = \frac{B}{2}\int_0^L (d\vec{t}/ds)^2 ds, \tag{2}$$

where DNA chain is fully described by a continuous chain (figure 1B) parameterized by $s$ with unit tangent vector $\vec{t}(s)$ and bending is characterized locally by the change of chain tangent $d\vec{t}/ds$. Correspondingly, the tangent vectors at different chain locations are correlated such that $\langle \vec{t}(s_1) \cdot \vec{t}(s_2) \rangle = e^{-|s_1-s_2|/l_P}$. Calculations based on this continuous description, using tools from path integral methods, show excellent agreement between theory and experiment for long DNA chains of $L \gg l_P$. Detailed variations of the WLC model have been proposed by introducing additional independent parameters, such as the twisting persistence length $l_T$[15].

## II. Model description and results

For a number of problems of great biological importance, the length scale of interest is similar to or even smaller than $l_P$. To improve the WLC model in describing DNA at short length scales[16], new models have been proposed with the introduction of new structural features, e.g. bubbles[17], kinks[18], or sub-elastic modes[19]. The general idea is to introduce additional terms in the bending energy, and thus provide new mechanisms to lower bending energy. Here we follow this spirit and consider a newly observed[20] and studied[21] feature of DNA: the correlation between local deformations. According to these recent studies, there exists a new length scale $l_D \sim 10$ bps, over which the local deformations of DNA—in particular, the major groove width deformations—are correlated. If we assume that local bending deformations behave in a similar manner, then the bending energy (eq. 1) considered in the KP model, where local bend angles are treated as independent, is only valid for $l_0 \gg l_D$. Since the WLC model as the continuous limit of the KP model is obtained at large $N$ limit, the WLC predictions only apply to sufficiently long DNA chains as $L = Nl_0$. For shorter chains of interest, the correlation between local bend angles needs to be incorporated into the definition of a more general bending energy. Here in our correlated WLC (C-WLC) model, we define a simple bending energy that considers only the correlation between neighboring bend angles with segment length $l_0 = l_D$:

$$E_{C-WLC} = \frac{B'}{2l_D} \sum_n (\theta_n^2 + C * \theta_n \theta_{n+1}), \tag{3}$$

where $B' = l'_P k_B T$ is the local bending modulus, which is shown to be related to $l_P$ later, and $C = e/(1 + e^2) \sim 0.324$ is the coupling strength. This more generally defined bending energy predicts a correlated distribution of $\{\theta_n\}$ as $\langle \theta_i \cdot \theta_j \rangle = \frac{2l_D(1+e^2)}{l'_P(1-e^2)} e^{-|i-j|}$. In the continuous limit, this yields not only a correlation for the tangents $\langle \vec{t}(s_1) \cdot \vec{t}(s_2) \rangle \sim e^{-|s_1-s_2|/l_P}$ for $|s_1 - s_2| \gg l_P$ but also an additional short-ranged correlation for the derivatives of the tangent vectors $\langle \frac{d\vec{t}}{ds}(s_1) \cdot \frac{d\vec{t}}{ds}(s_2) \rangle \sim e^{-|s_1-s_2|/l_D}$, which is absent in the WLC model.

In an approximate but simple way, the flexibility of a C-WLC can be quantitatively studied by mapping it to a WLC of the same contour length with an effective persistence length $l_{EP}$, so that earlier analytical results can apply[22, 23]. For a DNA chain of contour length $L = (N + 1)l_D$, the mapping can be done by matching the end-to-end tangent correlation obtained from the two models. If we limit our considerations up to near next neighbor correlations, in our C-WLC model the end-to-end tangent correlation is:

$$\langle \vec{t}_1 \cdot \vec{t}_{N+1} \rangle_{C-WLC} = (1 - \langle \theta_i^2 \rangle/2)^N * (1 - \langle \theta_i \theta_{i+1} \rangle)^{N-1} = e^{-Nl_D\left(1 - \frac{N-1}{N} 2e^{-1}\right)/l_P' \sqrt{1-4C^2}}. \qquad (4)$$

For a WLC with effective persistent length $l_{EP}$ the end-to-end tangent correlation is:

$$\langle \vec{t}(0) \cdot \vec{t}(Nl_D) \rangle_{WLC} = e^{-Nl_D/l_{EP}}. \qquad (5)$$

A comparison between eq. 4 and eq. 5 shows that the effective WLC has a contour length dependent persistence length

$$l_{EP}(L) = l_P' \sqrt{1 - 4C^2} \left(1 - \frac{N-1}{N} 2e^{-1}\right)^{-1} \xrightarrow{L \to \infty} \frac{l_P' \sqrt{1-4C^2}}{1 - 2e^{-1}}. \qquad (6)$$

From eq. 6 we see that the effective persistence length $l_{EP}(L)$ approaches its long chain limit quickly as $N = L/l_D - 1$ increases. At this long chain limit ($L \gg l_D \sim 10$ bps), our C-WLC model can be reduced to the WLC model by setting $l_{EP}(\infty) = l_P' \sqrt{1 - 4C^2}/(1 - 2e^{-1}) = l_P$. For a short chain, our model introduces a correction term and predicts a contour length dependent persistence length

$$l_{EP}(L) = l_P * \left(1 + \frac{2e^{-1}l_D}{(1-2e^{-1})L}\right)^{-1}, \qquad (7)$$

which is illustrated in figure 2 by setting $l_D = 10$ bps.

### III. Comparison to experimental measurements

The predictions from our C-WLC model may be compared to several recent experimental measurements. We focus on two very different sets of experimental studies, both of which serve as good tests of theoretical descriptions for DNA flexibility. In one set of experiments, the extension of a DNA chain is measured as a function of the amplitude of a pair of stretching forces $F$ at the two ends. At a contour length of about 100,000 bps[24], the experimentally observed force-extension behavior fits extremely well to the WLC prediction[5, 25] with a fitting parameter of $l_P = 53$ nm. In the other set of experimental studies, the probability of DNA forming a ring is measured, in terms of the J factor[26] defined as the ratio of equilibrium constants for cyclization and bimolecular association. Early J factor measurements have been compared to theoretical predictions, obtained by Shimada and Yamakawa in their seminal paper[22] where DNA is modeled as a Twisted WLC (TWLC). In this variation of the WLC model, the local orientation of DNA segment at $\vec{r}(s)$ is described by an orthonormal triad $\vec{e}_i(s)$ ($i = 1,2,3$), where $\vec{e}_3(s) = \vec{t}(s)$ is the tangent of the chain contour. The twist degree of freedom is considered explicitly by introducing the torsional energy $E_{Torsion} = 0.5 l_T k_B T \int_0^L (\omega_3 - \tau)^2 ds$, where $\omega_3(s) = \vec{\omega}(s) \cdot \vec{e}_3(s)$ is the local twist with $\vec{\omega}(s)$ determined from the relationship $d\vec{e}_i(s)/ds = \vec{\omega}(s) \times \vec{e}_i(s)$, and $\tau$ is the intrinsic twist. The J factor, $J_{TWLC}(L/l_P, \eta, \tau)$, is then obtained in terms of three parameters: $L/l_P$, $\eta \equiv l_P/l_T - 1$, and $\tau$. It has been shown[8] that the experimentally observed J factors at different contour lengths can fit well to this theoretical prediction with $l_P = 47$ nm, $\eta = -0.2$ and $\tau = 2\pi/10.5 = 0.6$. While both sets of experimental data support the WLC model, the difference in predicted persistence lengths should not be overlooked.

Here we attempt to resolve this difference and explain the short chain behavior using our C-WLC model. To highlight the effect of the new feature introduced in our model, for the J factor problem we minimize the number of free parameters by fixing $\eta = -0.2$ and $\tau = 0.6$.
Our C-WLC model then predicts the J factor as:

$$J_{C-WLC}(L) = J_{TWLC}(L/l_{EP}(L), -0.2, 0.6) = J_{TWLC}\left(\left(L + \frac{2e^{-1}l_D}{1-2e^{-1}}\right)/l_P, -0.2, 0.6\right). \quad (8)$$

Eq. 7 and eq. 8 show that, in addition to the persistence length $l_P$ for an infinitely long chain, our model derives another free parameter: the finite length correction $l_c = \frac{2e^{-1}l_D}{1-2e^{-1}}$. Here we fix $l_P = 53$ nm since it is obtained at long chain limit and fit the J factor measurements[27-29] to our theoretical predictions (eq. 8) with a single parameter—the finite length correction $l_c$. A good fit is obtained at $l_c = \frac{2e^{-1}l_D}{1-2e^{-1}} = 15$ bps and our theoretical predictions show a significant enhancement in terms of the J factor over the WLC predictions for short chains (figure 3A).

Despite that our model predicts a much higher flexibility for short DNA chains, it still falls short in explaining the recent experimental observations by Vafabakhsh and Ha[13], which reported a J factor about three orders of magnitudes larger than the classical results by Du et al.[27] at the same contour length of $L = 105$ bps (figure 3B). A comparison between the experimental methods shows that there might exist a capture radius for the study reported in ref. 13, that is, instead of measuring the ring closure probability, their results correspond to the probability with the two DNA ends separated by a distance $R$. Within the WLC framework, an analytical approximation for this probability, $Q_{WLC}(L/l_P, R/L)$, has been provided (eq. 21 in ref. 30), without consideration of the twist degree of freedom. By replacing $l_P$ with the contour length dependent persistence length $l_{EP}(L)$, the modified function $Q_{C-WLC}$ significantly enhances the resulting probability over the WLC result and shows qualitative agreement with the experimental measurements at a capture radius $R = 6$ nm (figure 3B).

The contribution of the twist degree of freedom can be characterized by the ratio of two theoretically evaluated J factors—$J_{TWLC}(L/l_P, \eta, \tau)$ and $J_{(1)}(L/l_P)$—obtained with and without consideration of DNA orientation alignment ($\vec{e}_i(s)$ for $i = 1,2,3$; or equivalently $\vec{\Omega}(s) = [\theta, \phi, \psi]$ the Euler angles), respectively. Evaluated from the analytical expressions[22], this ratio $P(L) \equiv J_{TWLC}(L/l_P, -0.2, 0.6)/J_{(1)}(L/l_P)$ fits well to a series of Gaussian functions separated by the helical pitch length $H = 10.5$ bps (figure 4A). This result suggests an approximate but simple treatment of the twist degree of freedom. In this treatment, we first assume that the number of microstates, in which the chain contour configuration $\{\vec{t}(s)\}$ deviates strongly from the most probable configuration $\{\vec{t}^*(s)\}$ (a circle of radius $L/2\pi$), is negligibly small. With the contour degree of freedom "frozen", we have $\omega_3(s) = d\psi(s)/ds$ and the total twist deformation needed for alignment

$$\gamma^*(L) = \int_0^L (\omega_3(s) - \tau)ds = \int_0^L (d\psi(s)/ds - \tau)ds = 2\pi(M - L/H), \qquad (9)$$

where $M$ is the closest integer to $L/H$. We can further assume that the behavior of the ratio, now modeled as $P(L) = 2\pi f(\gamma^*(L))/\int_0^{2\pi} f(\gamma)d\gamma$ with $f(\gamma)$ the probability associated with a total twist deformation of $\gamma$, is dominated by the most probable configuration in twist ($\omega_3(s) - \tau = const = \gamma^*/L$) so that $f(\gamma) = \frac{l_T}{2L}\left(\frac{2\pi}{H}\right)^2(L - MH)^2$. Then the ratio is further simplified as $P(L) = \frac{H}{\sqrt{2\pi}\sigma'}e^{-\frac{(L-MH)^2}{2\sigma'^2}}$, a series of Gaussian functions with standard deviation $\log_{10}\sigma'(L) = a\log_{10}L + b$ with $a = 0.5$ and $b = -0.900$. The standard deviations obtained from the analytical result of $P(L)$ as shown in figure 4A can fit to the same function form of $L$, with fitting parameters $a = 0.505$ and $b = -0.931$ (figure 4B). This close agreement supports the use of our simple treatment, which allows us to model the probabilities obtained experimentally in ref. 13 as $Q_{TC-WLC}(L,R) = Q_{C-WLC}(L,R) * P(L)$. To better explain the experimental data, a tolerance angle $\alpha = 0.9$ radian is introduced so that the configuration captured in experiment can allow an orientation shift with $|\psi(L) - \psi(0)| < \alpha$. Eq. 9 is modified accordingly with $\gamma^*(L) = 0$ when $|2\pi(M - L/H)| < \alpha$ and $\gamma^*(L) = [2\pi(M - L/H) - \alpha]\frac{2\pi(M-L/H)}{|2\pi(M-L/H)|}$ otherwise, and our results of $Q_{TC-WLC}(L, 6nm)$ show a clear oscillatory behavior as a function of $L$ and match the experimental measurements reasonably well (figure 3B).

## IV. Conclusion and discussion

The WLC model has been successful in describing the DNA flexibility at large length scales by introducing $l_P \sim 150$ bps to characterize the correlation between the tangents of the long chain. With a more generally defined bending energy (eq. 3), our C-WLC model improves the WLC model and extends the description of DNA flexibility to a shorter regime, by introducing a new length scale $l_D \sim 10$ bps to characterize a higher order correlation between the derivatives of the tangents. Our model shows analytically that there exists a finite length correction term $l_c$, leading to a contour length dependent persistence length (eq. 7), which approaches to the WLC prediction at infinite long chain limit. As a result, short DNA chains show notable "softening" as $l_c/L$ approaches 1, yielding a significant enhancement of the J factor, in agreement with recent experimental observations.

Our prediction of a finite length correction is derived using no assumptions about the boundary conditions, and is expected to apply to any segment of DNA in the middle of a long DNA chain with two ends $s_1$ and $s_2$ where $|s_2 - s_1| = L$. The generality of the derivation shows that this finite length correction is not an artifact of boundary conditions or a result of end effects, but a consequence of the cooperative bending behavior of DNA at local length scales not considered in the WLC model. While the details of the cooperative bending behavior as an intrinsic feature of DNA are still missing, the fitted value of the finite length correction $l_c = \frac{2e^{-1}l_D}{1-2e^{-1}} = 15$ bps provides an estimate of the correlation length scale $l_D \sim 5$ bps. This value is close to the length scale (~10 bps) over which the major groove width deformations are correlated, as observed in recent allosteric protein binding experiment[20], suggesting these cooperative behaviors may share the same origin. Further characterization of the cooperative behaviors will elucidate this connection.

**Figure captions**

**Figure 1. Description of worm-like chain model.** (**A**) Illustration of the Kratky-Porod model. (**B**) Illustration of the worm-like chain model.

**Figure 2. Contour length dependent persistence length.** The flexibility of correlated worm-like chain can be described by a persistence length $l_{EP}(L)$, which is illustrated as a function of the contour length $L$.

**Figure 3. Ring closure probability (J factor) as a function of the contour length.** (**A**) The experimental data taken from Du et al.[27] (red circles), Vologodskaia & Vologodskii[28] (red asterisks), and Shore & Baldwin[29] (red triangles) are compared with the WLC predictions (green solid line) and our C-WLC predictions (blue solid line). (**B**) The experimental data taken from Du et al.[27] (solid black circles) are compared with our C-WLC predictions $J_{C-WLC}$ when twist alignment is considered (blue solid line) and $J_{(1)}$ when twist alignment is not considered (blue dashed line). The experimental data for a finite capture radius $R$ taken from Vafabakhsh & Ha[13] (open black squares) are compared to the theoretical predictions of $Q_{WLC}$ (green dashed line), $Q_{C-WLC}$ (red dashed line), and $Q_{TC-WLC}$ (red solid line) at a capture radius $R = 6$ nm. DNA parameters used in both WLC model and C-WLC model include $l_P = 53$ nm, $\eta = -0.2$, $\tau = 0.6$, and helical pitch $H = 10.5$ bps $= 3.57$ nm, while C-WLC model has one more parameter, the finite length correction $l_c = 15$ bps.

**Figure 4. Contribution of the twist degree of freedom.** (**A**) The results (symbols) of the ratio $P(L) \equiv J_{TWLC}(L/l_P, -0.2, 0.6)/J_{(1)}(L/l_P)$, fit well to a Gaussian function (line) in each of the following color coded regions: 60bps $< L <$ 70bps (brown), 70bps $< L <$ 80bps (green),

80bps $< L <$ 90bps (blue), 90bps $< L <$ 100bps (red), and 100bps $< L <$ 110bps (black).

(**B**) The standard deviations and the corresponding center locations (black circles), obtained from the Gaussian fits in (**A**), fit well to a linear function (red line) with slope $a = 0.505$ and intercept $b = -0.931$ in a logarithm-logarithm plot.

**Acknowledgements**

X.L.X. and J.C. acknowledge the financial assistance of Singapore-MIT Alliance for Research and Technology (SMART), National Science Foundation (NSF CHE-112825), Department of Defense (DOD ARO W911 NF-09-0480), as well as a research fellowship by Singapore University of Technology and Design (to X.L.X.). The research work by B.J.R.T. is supported by Singapore University of Technology and Design Start-Up Research Grant (SRG EPD 2012 022).

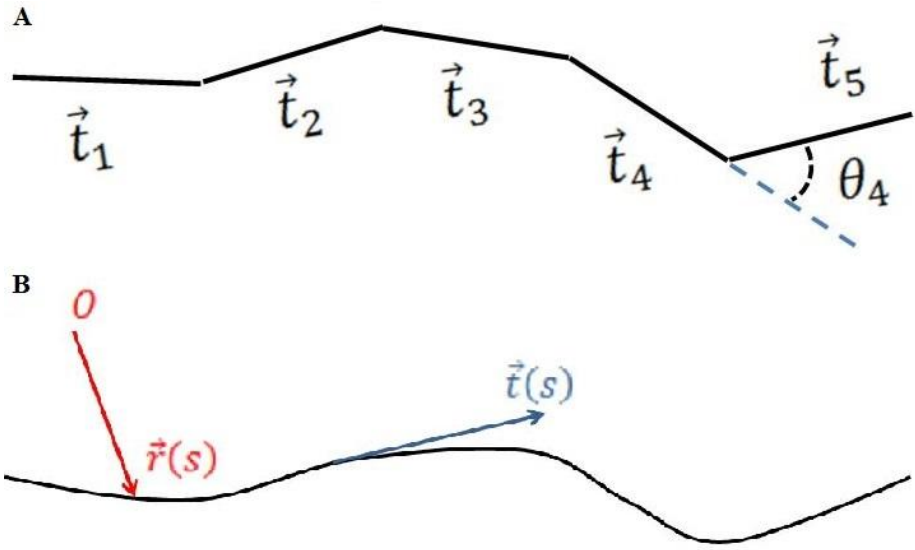

Figure 1
17

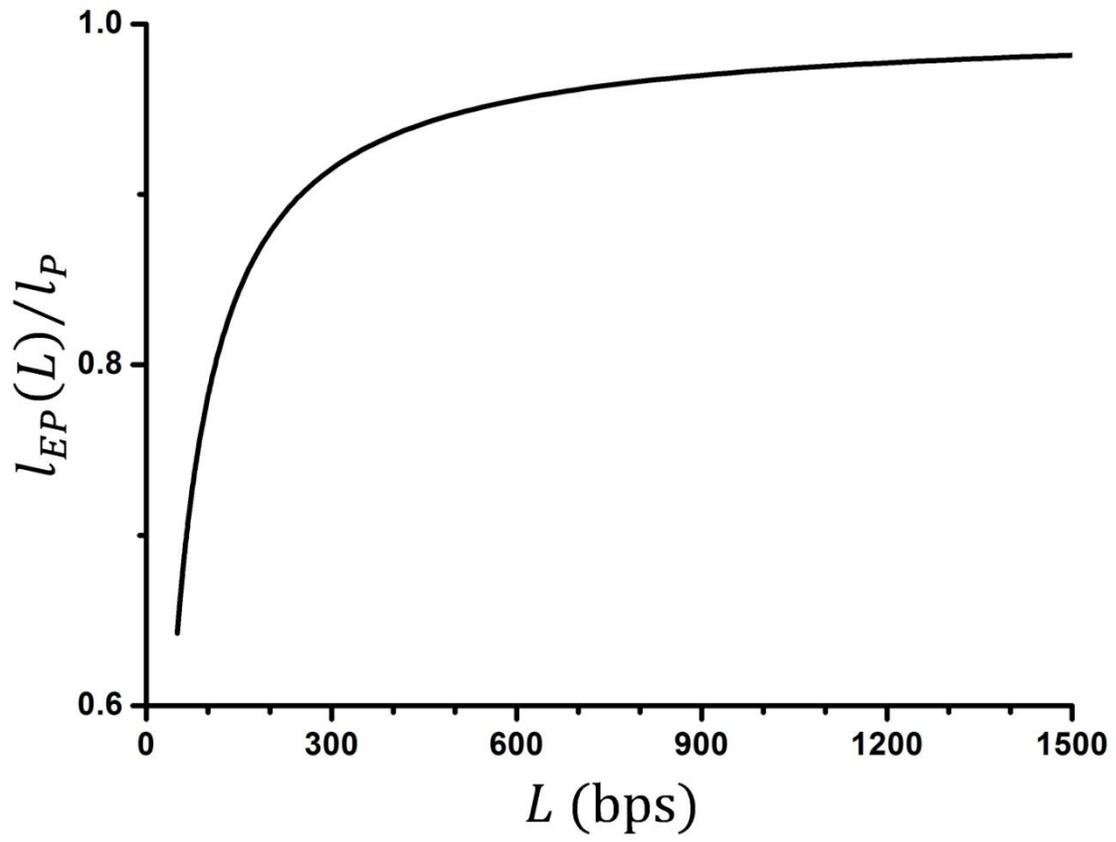

Figure 2.

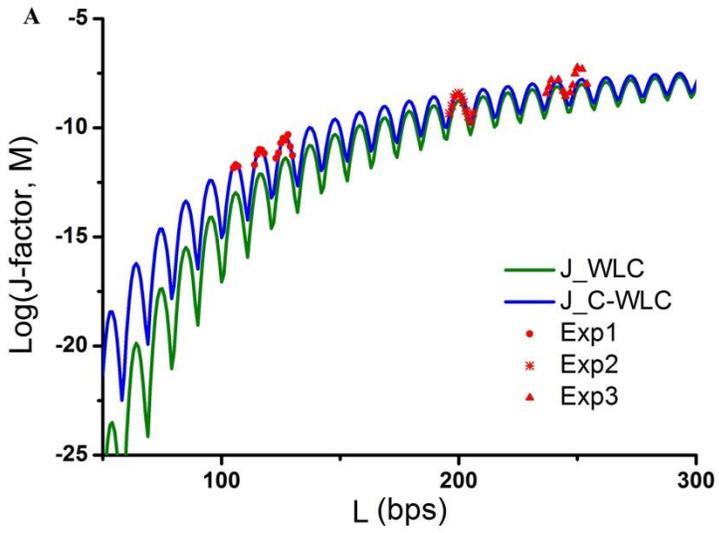
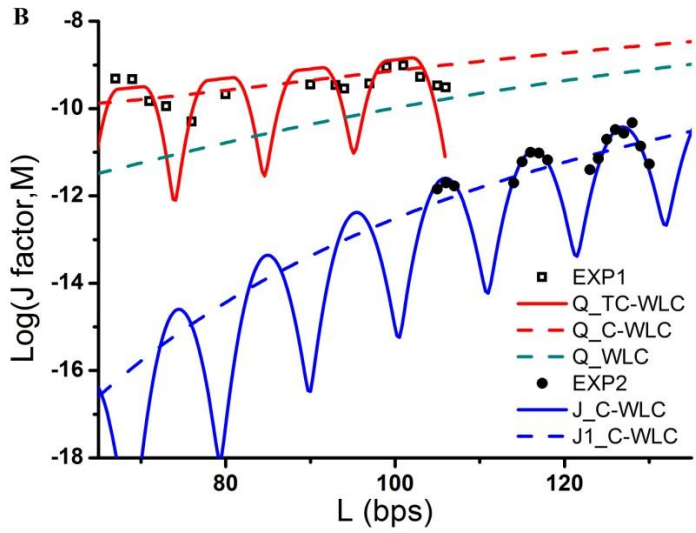

Figure 3.

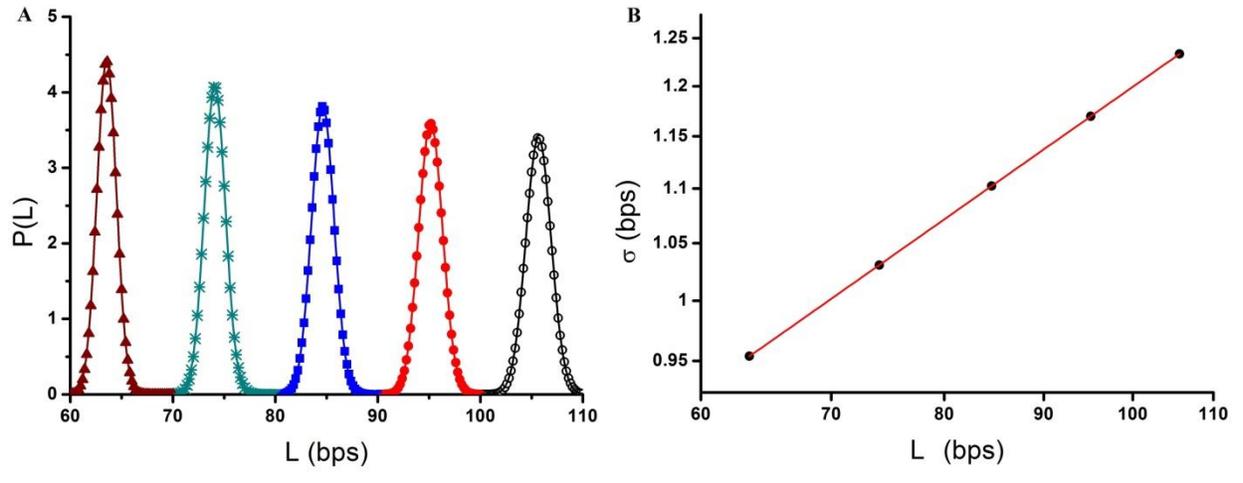

Figure 4.